\documentclass[%
 reprint,
 amsmath,amssymb,
 aps,
]{revtex4-1}

\usepackage{graphicx}
\usepackage{dcolumn}
\usepackage{bm}

\begin{document}

\preprint{APS/123-QED}

\title{Back-reaction of quantum processes and modified gravitational dynamics}

\author{Vyshnav Mohan}

 \email{vyshnav@ug.iisc.in}
\affiliation{%
 Indian Institute of Science, Bangalore 560012, India \\
 }%

\begin{abstract}
In this paper, we seek to find  a modified theory of gravity that accounts for the back-reaction of QED on curved spacetime. It is already known that vacuum fluctuations induce interactions between gravity and photons. An effective action for electromagnetism, which encodes the details of such quantum process, is utilized to get a set of modified Maxwell's equation and a new Lagrangian for the dynamics of gravity. Imposition of Cauchy predictability on these modified Maxwell's equations lead us to a new effective metric, from which the Lagrangian can be calculated. The new Lagrangian for gravity turns out to be a function of not just the higher order derivatives of the metric tensor but also the polarization of the photon. This immediately results in phenomenon such as gravitational birefringence and existence of black holes with polarization dependent event horizons.

\end{abstract}

\pacs{Valid PACS appear here}
\maketitle


\section{\label{sec:level1}Introduction}
 Quantum electrodynamics is one of the most successful theories of theoretical physics and has made many significant predictions about the way our universe behaves. One of the consequences of QED is that it provides corrections to local photon propagation on a curved spacetime. This effect, produced by vacuum polarization, on the velocity of light and the light cone structure has already been analyzed in various background spacetimes by Drummond and Hathrell  \cite{drum80}. Working in one-loop approximation, they found out an effective action for the theory. We can physically motivate this modification as the vacuum polarization are processes in which the photon exists as a virtual electron and positron pair and these species behave differently from photons in a general curved spacetime. The same line of thought can be seen in the works of Klaus Scharnhorst and others like Sergei Odintsov. \cite{Scharn93,odin92,Odintsov:1990mt}

  However, if we want these modified matter equations to be Cauchy predictable, the dynamics of the underlying gravitational field becomes constrained as shown in Ref. \cite{fps14, Rat10, kris12}. This is because suitable initial data surface are required to uniquely evolve the initial data. It is imperative that a classical theory must be predictable and thus, the constraints on gravity arise naturally. It turns out that the geometric metric tensor ceases to determine the dynamics of the gravitational field and some quantum corrections are required to fix this issue. The details of these corrections are worked out in the next section.

 Thus, the predictability condition helps us to assess the influence of QED vacuum on the gravitational field.  This process of finding a modified theory of gravity is quite analogous to that of finding the electric field inside a sphere of linear dielectric material in classical electrodynamics. Assume that there is a uniform static electric field and the sphere is embedded in it. The field polarizes the sphere and produces bound charges. These charges in turn produce a field of its own and impart more polarization to the sphere. This process continues and each iteration further improves the accuracy of the field calculated inside the object and one can see in Ref. \cite{grif12} that it converges to the conventional value obtained by other methods. 

Analogously, we start off by putting quantum fields on a static background spacetime and look at the modification of the matter equations and use this information to find the corrections to the background gravitational field. The details of these calculations are given in the rest of the paper.

\section{ The effective metric}
 The Drummond and Hathrell effective action $\Gamma_{\text{DH}}$ can be calculated by accounting for one-loop corrections in QED \cite{drum80, shore03}  i.e.
\begin{equation}
\Gamma_{\text{DH}} = \Gamma_{M} +\text{ln det S}(x, x^{'})                                                                                                                                                                                                                                                                                                                                                                                                                                                                                                                                                                                                                                                                                                                                                                                                                                                                                                                                                                                                                                                                                                                                                                                                                                                                                                                                      
\end{equation}

Here, $\Gamma_{M}$ is the free Maxwell action and S$(x, x^{'})$ is the electron propagator on the curved spacetime. Suppressing higher order derivatives of the curvature tensors relative to $O({\lambda}/{L})$, where $\lambda$ is the photon wavelength and $L$ is the typical curvature scale, we can find out the effective action as in Ref.\cite{drum80}.
After varying the action, we obtain the following modified Maxwell's equation \cite{shore03}:
\begin{equation}
D_{\mu} F^{\mu \nu} -\dfrac{1}{m^2} [2\sigma R_{\mu \nu}D^{\mu} F^{\lambda \nu} + 4 \zeta g^{ \nu \kappa} R_{\mu \kappa \lambda \rho}D^{\mu} F^{\lambda \rho}]  \label{eq:three}
\end{equation}
Here, $ \sigma =\frac{13\alpha}{360\pi}$ , $\zeta=-\frac{ \alpha}{360\pi}$ where $\alpha$ is the fine structure constant and $m$ is the mass of the electron. All the physical quantities are in natural units and $ F^{\mu \nu}$ is the electromagnetic field tensor as usual.

In order to focus on the characteristics of photon propagation, we use geometric optics approximation,where the electromagnetic field tensor is given by a product of  slowly varying amplitude and rapidly varying phase as \cite{sch92,shore03}
\begin{equation}
 F_{\mu \nu} =  f_{\mu \nu} e^{i \theta}
\end{equation}

 and the wave vector is given by $k_{\mu} = \partial_{\mu} \theta$. Together with Bianchi identities $D_{[\mu} F_{\lambda \nu]}=0 $, we get the following constraint
\begin{equation}
  f_{\mu \nu} =  k_{\mu}a_{\nu} -  k_{\nu}a_{\mu}
\end{equation}
 Where $ a^{\nu} $ is a vector in the direction of the polarization of the photon and can be assumed to obey the relation $k_{\mu}a^{\mu}= 0$.
Thus,  Eq.~(\ref{eq:three}) gives the modified light cone condition as in \cite{sch92,shore03,shore96} given by
\begin{equation}
 { g_{ab} k^{a} k^{b} +  {\dfrac{2 \sigma }{m^2}} \hspace{.09cm}   R_{ab} k^{a} k^{b}-  {\dfrac{8 \zeta}{m^2}} \hspace{.09cm}   R_{acbd}k^{a} k^{b}a^{c} a^{d} =0 }.
\end{equation}
Here, the polarization 4-vector is assumed to be space like normalized with respect to $g_{ab}$ and the latin indices run from $ 0$ to $3$. 

\begin{equation}
{ \text{Let}  \hspace{.09cm}  G_{ab} := g_{ab} +  {\dfrac{2 \sigma}{m^2}} \hspace{.09cm}   R_{ab}-  {\dfrac{8 \zeta}{m^2}}  \hspace{.09cm}  R_{acbd}a^{c} a^{d} }
\end{equation}
Thus, the modified light cone condition can be written as 
\begin{equation}
  G_{ab}  k^{a} k^{b} = 0
\end{equation}
This is strikingly similar to the light cone condition in general relativity, given by:
\begin{equation}
  g_{ab}  k^{a} k^{b} = 0
\end{equation}
 The predictability of the matter equations translates to the condition that the principle polynomial of the matter field should be a bi-hyperbolic and energy distinguishing homogeneous polynomial [See Ref.\cite{Rat10,fps14} for detailed discussions]. From the light cone condition or from the coefficients of the highest-order derivative in Eq.~(\ref{eq:three}), we can read off the dual polynomial for the system as
\begin{equation}
 {  P^{\#} (x,v) = G_{ab} v^{a} v^{b} }.
\end{equation}
Certain properties of $G_{ab}$ can easily be seen from the definition of the tensor. Due to the symmetries of Ricci and Riemann-Christoffel tensor, we find that $G_{ab}$ is a billinear, symmetric, non- degenerate map on that tangent space $T_{x}M$ of the manifold $M$ at every point $x$. Thus, $G_{ab}$ by itself can be treated as an effective metric of the manifold, provided an inverse metric exist at all points on the spacetime. It turns out that it is even possible to define $G _{ab}$ as a metric tensor of the manifold and find appropriate $g_{ab}$ which solves for $G _{ab}$ and satisfies its the non-degeneracy condition. 

Thus, we calculate the principal polynomial to be
\begin{equation}
 {  P (x,k) = G^{ab} k_{a} k_{b} }.
\end{equation}
 where $ G^{ab} $ is the inverse of the metric $ G_{ab} $. Another compelling reason to consider $ G_{ab} $ as an effective metric is the following:

 The wave vector $k_{\mu}$ is defined as a one-form belonging to $T_x^{\ast}(M)$  at every point of spacetime while the momentum $p^{\nu}$ of the photon belongs to the tangent space $T_{x}(M)$. In this modified theory, the relationship between them is not trivial as  $ p^{\nu} = g^{\mu \nu} k_{\mu}$. Instead the wave vector is mapped to the momenta up to a factor by the Gauss map (See Ref.\cite{Rat10}) given by $[k] \mapsto \bigl[\dfrac{\partial P}{\partial k_{a}}(x,k)\bigl] $ where $[X]$ represents the projective equivalence class of all vectors collinear with X. Thus, we see that the correct relationship between $ p^{\nu}$ and  $k_{\mu}$ is:
\begin{equation}
p^{a} =  \dfrac{\partial ( G^{ab} k_{a} k_{b})}{\partial k_{a}} = G^{ab} k_{b}
\end{equation}
 This shows that $G^{ab}$ takes the role of $g^{ab}$ in raising and lowering the indices of vectors and co-vectors. Thus, $G^{ab}$ can be treated as an effective metric to the manifold.

For the modified Maxwell's equations to have a well posed initial value problem, the principal polynomial should be hyperbolic. Since the underlying geometry is determined by an effective metric $ G^{ab} $, the hyperbolicity of dual polynomial reduces to the condition that there exist some vector field $h$ such that $ P^{\#}(h)  >0$ and for any vector field q, the following equation has only two real roots
\begin{equation}
 {  P^{\#} (x,h + \lambda q) =0 }.
\end{equation}
 Expanding the equation in terms of $ G_{ab} $ results in a quadratic equation of $ \lambda $ as in Ref.\cite{fps14}. Thus, the condition boils down to the discriminant of the quadratic equation being positive i.e.
\begin{equation}
 { G_{ab} h^{a} q^{b} -  G_{ab} h^{a} h^{b }  \hspace{.09cm}G_{cd} q^{c} q^{d} >0 }.
\end{equation} 

  Let us choose a vector basis \{$\epsilon_{\alpha}$\} such that $\epsilon_{0} := h $ and $ G_{ab} \hspace{.09cm}  \epsilon^{a}_{0} \epsilon^{b }_{\alpha} = 0 $

As $ G_{ab}\epsilon^{a}_{0}\epsilon^{b }_{0} > 0$, the discriminant becomes  $ q^{\alpha} q^{\beta} \hspace{.09cm}  G_{ab}  \hspace{.09cm} \epsilon^{a}_{\alpha} \epsilon^{b }_{\beta}< 0 $.
As $ g_{ab}$ is lorentzian, we find that if $ g_{ab} h^{a} h^{b} >0$ , then $ q^{\alpha} q^{\beta} \hspace{.09cm}  g_{ab}  \hspace{.09cm} \epsilon^{a}_{\alpha} \epsilon^{b }_{\beta}< 0 $
\begin{equation}
 \Rightarrow {\dfrac{2 \sigma}{m^2}}  \hspace{.09cm}  R_{\alpha \beta} -  {\dfrac{8 \zeta}{m^2}} \hspace{.09cm}   R_{\alpha c \beta d} \hspace{.09cm} a^{c} a^{d} < 0 \label{eq:one}
\end{equation}

\begin{equation}
 {\dfrac{2 \sigma}{m^2}} \hspace{.09cm}   R_{00}-  {\dfrac{8 \zeta}{m^2}} \hspace{.09cm}   R_{0 c 0d} \hspace{.09cm} a^{c} a^{d} > 0 \label{eq:two}
\end{equation}
 Here $\alpha,  \beta$ runs from 1 to 3 and   $R_{\alpha \beta},  R_{\alpha c \beta d}$ are the components of the tensors w.r.t basis  \{$\epsilon_{\alpha}$\}

Multiplying inequality Eq.~(\ref{eq:one}) by  $g^{\alpha \beta}$ and multiplying inequality Eq.~(\ref{eq:two}) by  $g^{00}$  summing up the indices, we get,
\begin{equation}
{\dfrac{2 \sigma}{m^2}}  g^{\alpha \beta} \hspace{.09cm}  R_{\alpha \beta} -  {\dfrac{8 \zeta}{m^2}} g^{\alpha \beta} \hspace{.09cm}   R_{\alpha c \beta d} \hspace{.09cm} a^{c} a^{d} > 0
\end{equation}
\begin{equation}
  {\dfrac{2 \sigma}{m^2}} g^{00} \hspace{.09cm}   R_{00}-  {\dfrac{8 \zeta}{m^2}} g^{00} \hspace{.09cm}   R_{0 c 0d} \hspace{.09cm} a^{c} a^{d} > 0
\end{equation}
 These inequalities are satisfied if 
\begin{equation}
 \hspace{.09cm} {\dfrac{2 \sigma}{m^2}}  \hspace{.09cm}   R > {\dfrac{8 \zeta}{m^2}} \hspace{.09cm}   R_{ c d} \hspace{.09cm} a^{c} a^{d}  \end{equation}

Conversely, if inequality (3) is satisfied along with the condition that the metric  $g_{ab}$ is lorentzian, the matter equations will have  a well posed initial value problem.

\section{ The Lagrangian}
Obtaining the Lagrangian from the principal polynomial is quite tedious [3]. However, as the geometry of the spacetime is described by a metric $G_{ab}$ ,we can use the hindsight from Einstein-Hilbert action that the gravitational part of the new Lagrangian must contain a term proportional to the Ricci scalar of the metric $G_{ab}$.
\begin{equation}
 \therefore \mathcal{L}= \int  \left[ {\dfrac{1}{2\kappa}} \left( \mathcal{R} - 2 \Lambda \right) + \mathcal{L}_\mathrm{M} \right] \sqrt{-g}\, \mathrm{d}^4 x   \hspace{.09cm},  \text{where}  \hspace{.09cm}{\large{\kappa = 8 \pi G}}
\end{equation}
$\mathcal{R}$ is the Ricci tensor obtained from the metric $ G_{ab}$, $\Lambda$ is the cosmological constant and $ \mathcal{L}_\mathrm{M}$ is the Lagrangian of any matter fields appearing in the theory.

$\mathcal{R}$ contains the Ricci scalar of the metric  $g_{ab}$ and higher order derivatives of Ricci and Riemann--Christoffel tensor of the  metric  $g_{ab}$ .

\section{Modified Einstein's equation} 

 The equation of motion of the system can be obtained by varying the Lagrangian w.r.t the metric $g_{ab}$. As the $\mathcal{R}$ term is messy, it is difficult to approach the problem directly. So, we vary the Lagrangian with  $G_{ab}$ and then change the variation from $G_{ab}$ to $g_{ab}$. 

Focusing on the gravitational part of the Lagrangian and varying w.r.t to $G_{ab}$, we get

\begin{equation}
{  { \hspace{.2cm} \int 
        \left[ 
           {\dfrac{1}{2\kappa} } \left( \dfrac{\delta R}{\delta G_{ab}} +
             \dfrac{R}{\sqrt{-G}} \dfrac{\delta \sqrt{-G}}{\delta G_{ab} } 
            \right)  
        \right] \delta G_{ab} \sqrt{-G}\, \mathrm{d}^4x}} \hspace{1.5cm} 
\end{equation}
However, from the definition of $G_{ab}$. we have 

\begin{equation}
  \delta G_{ab} =  \delta g_{ab} +  {\dfrac{2 \sigma}{m^2}} \hspace{.09cm}   \delta R_{ab}-  {\dfrac{8 \zeta}{m^2}}  \hspace{.09cm}   \delta R_{acbd}a^{c} a^{d}          
\end{equation}

We analyze each term separately.
We know that from Ref. \cite{wein72}
\begin{equation}
\delta R^h{}_{cld} =\nabla_l (\delta \Gamma^h_{dc}) - \nabla_d (\delta \Gamma^h_{lc})
\end{equation}
\begin{equation}
 \text{and}\hspace{.2cm} \delta {\Gamma^{\alpha}}_{\beta \mu}=  \dfrac 12 g^{\alpha \lambda} (\nabla_{\beta} \delta g_{\lambda \mu}+ \nabla_{\mu} \delta g_{\beta \lambda}-
\nabla_{\lambda} \delta g_{\beta \mu})  
\end{equation}
\begin{eqnarray}
\therefore  \hspace{.09cm} \delta R^h{}_{cld} =\nabla_l ( \dfrac 12 g^{he} (\nabla_{d} \delta g_{ec}+ \nabla_{c} \delta g_{de}-
\nabla_{e} \delta g_{dc}) ) \nonumber \\
 - \nabla_d ( \dfrac 12 g^{he} (\nabla_{l} \delta g_{ec}+ \nabla_{c} \delta g_{le}-
\nabla_{e} \delta g_{lc}) )~.
\end{eqnarray}
\begin{eqnarray}
 \hspace{1.6cm} = \dfrac 12 g^{he} (\nabla_l \nabla_{d} \delta g_{ec}+\nabla_l  \nabla_{c} \delta g_{de}-
\nabla_l \nabla_{e} \delta g_{dc} \nonumber \\
 - \nabla_d \nabla_{l} \delta g_{ec}+ \nabla_d \nabla_{c} \delta g_{le}-
\nabla_d  \nabla_{e} \delta g_{lc}  )~ ~ .
\end{eqnarray}
\begin{equation}
\text{ And} \hspace{0.09cm}   \delta R_{acbd} = \delta (g_{kh}   R^h{}_{cld}) =  \delta g_{kh}   R^h{}_{cld}+   g_{kh}  \delta  R^h{}_{cld}  , 
\end{equation}

From pallatani identities [8], we have
\begin{eqnarray}
\delta R_{kl} =\dfrac 12 g^{ab} (\nabla_a \nabla_{l} \delta g_{kb} +\nabla_a \nabla_{k} \delta g_{lb} \nonumber \\
-\nabla_k \nabla_{l} \delta g_{ab} - \nabla_a \nabla_{b} \delta g_{kl}  ) ~ .
\end{eqnarray}

Therefore the variation can be written as
\begin{eqnarray}
\delta \mathcal{L} = { \int  {1 \over 2\kappa}( \mathcal{G}^{kl} \hspace{.09cm} \delta g_{kl} +  \mathcal{G}^{kl} \hspace{.09cm} {\dfrac{2 \sigma}{m^2}} \hspace{.09cm}   \delta R_{kl}} \nonumber \\
 -  \mathcal{G}^{kl} \hspace{.09cm} {\dfrac{8 \zeta}{m^2}}  \hspace{.09cm}   \delta R_{kcld}a^{c} a^{d}) \hspace{.09cm} \sqrt{-G}\,\hspace{.09cm}\mathrm{d}^4x  ~ .
\end{eqnarray} 

 where $ \mathcal{G}^{kl}$ is the Einstein tensor associated with metric $G_{ab}$. Consider the second term of the integral.

 $ { \int {1 \over 2\kappa}\mathcal{G}^{kl}{\dfrac{2 \sigma}{m^2}} \hspace{.09cm}   \delta R_{kl}  \sqrt{-G}\,\hspace{.09cm} \mathrm{d}^4x} $ 

\begin{eqnarray}
= \int {1 \over 2\kappa}\mathcal{G}^{kl} \hspace{.09cm} {\dfrac{2 \sigma}{m^2}} \hspace{.09cm} ( \dfrac 12 g^{ab} (\nabla_a \nabla_{l} \delta g_{kb}+\nabla_a \nabla_{k} \delta g_{lb} \nonumber \\
 -\nabla_k \nabla_{l} \delta g_{ab} - \nabla_a \nabla_{b} \delta g_{kl}  ) ) \hspace{.09cm} \sqrt{-G}\,\hspace{.09cm} \mathrm{d}^4x  ~ .
\end{eqnarray} 

Now, we use integration by parts twice to shift the covariant derivative from the variation of the metric and modify the dummy indices to get :
\begin{flushleft}
\hspace{1cm} $ { \int {1 \over 2\kappa} {\dfrac{2 \sigma}{m^2}} \hspace{.09cm} \nabla_b \nabla_{a}( g^{al} \hspace{.09cm} \mathcal{G}^{bk} - \dfrac 12 g^{kl} \hspace{.09cm} \mathcal{G}^{ab}- \dfrac 12 g^{ab} \hspace{.09cm} \mathcal{G}^{kl}) \hspace{.09cm} \delta g_{kl} \hspace{.09cm} \sqrt{-G}\,\hspace{.09cm} \mathrm{d}^4x}$ .
\end{flushleft}
 Now consider the third term
\begin{flushleft}
$ { \int {1 \over 2\kappa}\mathcal{G}^{kl} \hspace{.09cm} {\dfrac{8 \zeta}{m^2}}  \hspace{.09cm}   \delta R_{kcld}a^{c} a^{d}  \sqrt{-G}\,\hspace{.09cm} \mathrm{d}^4x} $ 
$= { \int {1 \over 2\kappa}\mathcal{G}^{kl} \hspace{.09cm} {\dfrac{8 \zeta}{m^2}}  \hspace{.09cm}  (  \delta g_{kh}   R^h{}_{cld}+   g_{kh}  \delta  R^h{}_{cld} )a^{c} a^{d}  \sqrt{-G}\,\hspace{.09cm} \mathrm{d}^4x} $
\end{flushleft}

Let us focus on the second term of the above expression i.e.

$ { \int {1 \over 2\kappa}\mathcal{G}^{kl} \hspace{.09cm} {\dfrac{8 \zeta}{m^2}}  \hspace{.09cm}  g_{kh}  \delta  R^h{}_{cld} a^{c} a^{d}  \sqrt{-G}\,\hspace{.09cm} \mathrm{d}^4x} \nonumber $

\begin{eqnarray}
=  \int {1 \over 2\kappa}\mathcal{G}^{kl} \hspace{.09cm} {\dfrac{8 \zeta}{m^2}} \hspace{.09cm} g_{kh} ( \dfrac 12 g^{he} (\nabla_l \nabla_{d} \delta g_{ec}+\nabla_l \nabla_{c} \delta g_{de}\nonumber \\
 -\nabla_l  \nabla_{e}  \delta g_{dc} -  \nabla_d \nabla_{l} \delta g_{ec}\nonumber \\
+ \nabla_d \nabla_{c} \delta g_{le} -\nabla_d \nabla_{e} \delta g_{lc} )) a^{c} a^{d} \sqrt{-G}\,\hspace{.09cm} \mathrm{d}^4x . ~ ~ ~ 
\end{eqnarray} 

Using the identity $\delta g_{ec}{}_{;ld}-\delta g_{ec}{}_{;dl}=R^a{}_{eld}\delta g_{ac} + R^a{}_{cld}\delta g_{ae}$ for first and fourth terms and using integration by parts twice on other terms and then changing the dummy indices, we get

${ \int {1 \over 2\kappa}\mathcal{G}^{kl} \hspace{.09cm} {\dfrac{8 \zeta}{m^2}}  \hspace{.09cm}  g_{kh}  \delta  R^h{}_{cld} a^{c} a^{d}  \sqrt{-G}\,\hspace{.09cm} \mathrm{d}^4x} $ 
\begin{eqnarray}
=  \int {1 \over 2\kappa}(\mathcal{G}^{ac} \hspace{.09cm} {\dfrac{4 \zeta}{m^2}}  \hspace{.09cm}  R^k{}_{acd} a^{l} a^{d} + \mathcal{G}^{la} \hspace{.09cm} {\dfrac{4 \zeta}{m^2}}  \hspace{.09cm}  R^k{}_{cad} a^{c} a^{d}\nonumber \\
 - \nabla_d \nabla_{c}( \mathcal{G}^{dc} \hspace{.09cm} {\dfrac{4 \zeta}{m^2}}  \hspace{.09cm}a^{l} a^{k})+\nabla_c \nabla_{d}( \mathcal{G}^{ld} \hspace{.09cm} {\dfrac{4 \zeta}{m^2}  \hspace{.09cm}a^{c} a^{k}) ) }\nonumber \\
   - \nabla_c \nabla_{d}( \mathcal{G}^{ck} \hspace{.09cm} {\dfrac{4 \zeta}{m^2}}  \hspace{.09cm}a^{l} a^{d})+\nabla_c \nabla_{d}( \mathcal{G}^{kl} \hspace{.09cm} {\dfrac{4 \zeta}{m^2}}  \hspace{.09cm}a^{c} a^{d}) ) \nonumber \\
  {\delta g_{kl} \sqrt{-G}\,\hspace{.09cm} \mathrm{d}^4x}) . ~ ~ 
\end{eqnarray} 

\begin{widetext}

$ \therefore$ The total variation is
\begin{center}
   $  \delta \mathcal{L} = { \int {1 \over 2\kappa}( \mathcal{G}^{kl} \hspace{.09cm} + {\dfrac{2 \sigma}{m^2}} \hspace{.09cm} \nabla_b \nabla_{a}( g^{al} \hspace{.09cm} \mathcal{G}^{bk}}- \dfrac 12 g^{kl} \hspace{.09cm} \mathcal{G}^{ab}- \dfrac 12 g^{ab} \hspace{.09cm} \mathcal{G}^{kl}) \hspace{.09cm} $
\end{center}
 \begin{center}
$ - \mathcal{G}^{ac} \hspace{.09cm}{\dfrac{4 \zeta}{m^2}} \hspace{.09cm}  R^k{}_{acd} a^{l} a^{d}  - \mathcal{G}^{la} \hspace{.09cm} {\dfrac{4 \zeta}{m^2}}  \hspace{.09cm}  R^k{}_{cad} a^{c} a^{d}  - \mathcal{G}^{kh} \hspace{.09cm} {\dfrac{8 \zeta}{m^2}}  \hspace{.09cm}  R^l{}_{chd} a^{c} a^{d}$
\end{center}
\begin{center}
 $- \nabla_d \nabla_{c}( \mathcal{G}^{dc} \hspace{.09cm} {\dfrac{4 \zeta}{m^2}}  \hspace{.09cm}a^{l} a^{k}) +\nabla_c \nabla_{d}( \mathcal{G}^{ld} \hspace{.09cm} {\dfrac{4 \zeta}{m^2}}  \hspace{.09cm}a^{c} a^{k} )$
\end{center}
\begin{center}
${+ \nabla_c \nabla_{d}( \mathcal{G}^{ck} \hspace{.09cm} {\dfrac{4 \zeta}{m^2}}  \hspace{.09cm}a^{l} a^{d}) -\nabla_c \nabla_{d}( \mathcal{G}^{kl} \hspace{.09cm} {\dfrac{4 \zeta}{m^2}}  \hspace{.09cm}a^{c} a^{d}) ) {\delta g_{kl} \sqrt{-G}\,\hspace{.09cm} \mathrm{d}^4x} } $ .
\end{center}
\begin{equation}
\stackrel{!}{=} 0 \hspace{.09cm} \text{( By principle of least action.)}
\end{equation}
 Thus, the integrand must vanish identically at all points. Therefore, the modified Einstein's equation is 

\begin{equation}
\mathcal{G}^{kl} \hspace{.09cm} + {\dfrac{2 \sigma}{m^2}} \hspace{.09cm} \nabla_b \nabla_{a}( g^{al} \hspace{.09cm} \mathcal{G}^{bk} -  \dfrac 12 g^{kl} \hspace{.09cm} \mathcal{G}^{ab}- \dfrac 12 g^{ab} \hspace{.09cm} \mathcal{G}^{kl})  - \mathcal{G}^{ac} \hspace{.09cm}{\dfrac{4 \zeta}{m^2}}  \hspace{.09cm}  R^k{}_{acd} a^{l} a^{d} - \mathcal{G}^{la} \hspace{.09cm} {\dfrac{4  \zeta}{m^2}}  \hspace{.09cm}  R^k{}_{cad} a^{c} a^{d} \nonumber
\end{equation}

\begin{equation}
 { - \mathcal{G}^{kh} \hspace{.09cm} {\dfrac{8 \zeta}{m^2}}  \hspace{.09cm}  R^l{}_{chd} a^{c} a^{d}  - \nabla_d \nabla_{c}( \mathcal{G}^{dc} \hspace{.09cm} {\dfrac{4 \zeta}{m^2}}  \hspace{.09cm}a^{l} a^{k}) +\nabla_c \nabla_{d}( \mathcal{G}^{ld} \hspace{.09cm} {\dfrac{4 \zeta}{m^2}}  \hspace{.09cm}a^{c} a^{k} )} \nonumber
\end{equation}

\begin{equation}
 {+ \nabla_c \nabla_{d}( \mathcal{G}^{ck} \hspace{.09cm} {\dfrac{4 \zeta}{m^2}}  \hspace{.09cm}a^{l} a^{d}) -\nabla_c \nabla_{d}( \mathcal{G}^{kl} \hspace{.09cm} {\dfrac{4 \zeta}{m^2}}  \hspace{.09cm}a^{c} a^{d})  = \kappa T^{kl}  }
\end{equation}
Where the $  T^ {kl}$ is the stress-energy tensor.
\end{widetext}
\section{Corrections to Schwarzschild metric }
  Solving directly for a homogeneous, static and spherically symmetric spacetime turns out to be a very difficult problem.Therefore, we linearize the metric and obtain the solutions to the modified equations. Thus, the metric is assumed to take the form 
\begin{equation}
 g = B(r) dt\otimes  dt+  A(r) dr\otimes  dr +  r^2( d\theta \otimes  d\theta +sin^2\theta  \hspace{.09cm} d\phi\otimes  d\phi )
\end{equation}
where $B(r) = -(1 +k (r)) $ and $ A(r) = 1 +j (r)$. Here  $k (r)$ and $j (r)$ are assumed to be infinitesimal in magnitude and we neglect their higher powers . We assume this applies to the higher order derivatives of the functions as well. In this way, we neglect the product or products of the function with their higher derivatives.This implies that the inverses are approximately given by
\begin{equation}
B ^{-1} (r) = -1 + k (r) \hspace{.09cm} \text{and}\hspace{.09cm}  A ^{-1} (r) = 1 -j (r)
\end{equation}
 However, upon close inspection of the modified Einstein's equations, we find that these differential equations are of the order of six in terms of  $k(r)$ and $j(r)$.  But we can bring the order of the equations down by using the fact that the we keep only terms that are linear in $k(r)$ and $j(r)$  and their derivatives.

 The calculation of the components of Riemann-Christoeffel tensor shows that all the non-vanishing terms are of infinitesimal magnitude. Hence, we can neglect the terms like  ${ \mathcal{G}^{kh} \hspace{.09cm} R^l{}_{chd} a^{c} a^{d}}$ as $ \mathcal{G}^{kh}  $ is also an infinitesimal quantity.
 Thus, the final equations comprise of terms containing $ \mathcal{G}^{kh}$  and $ a^{d}$ and their covariant derivatives.
 This motivates to us to start from $ G_{ab}$ rather than $ g_{ab}$. So , we define 
\begin{equation}
 G = \mathcal{B}(r) dt\otimes  dt+ \mathcal{A}(r) dr\otimes  dr \nonumber
\end{equation}
\begin{equation}
 \hspace{0.95cm} +  \mathcal{R}(r)^2( d\theta \otimes  d\theta +sin^2\theta  \hspace{.09cm} d\phi\otimes  d\phi )
\end{equation}

The component functions $\mathcal{B}(r)$, $\mathcal{A}(r)$ and $\mathcal{R}(r)$ can be calculated from the definition of $ G_{ab}$ in terms of $k (r)$ and $j (r)$, but instead we define it as the following 
\begin{equation}
\mathcal{B}(r) = 1 + v (r)  , \hspace{.09cm}  \mathcal{A} (r) = 1 +w (r) \label{metric}
\end{equation}
\begin{equation}
 \mathcal{R} (r) =r^2 +h (r)  
\end{equation}

and solve for $v (r)$, $w (r)$ and $h (r)$.

In the case of Schwarzschild solution for Einstein's field equations, we have the vanishing of all the components of the Einstein tensors as $  T^ {kl}$ is zero. Using this information and by inspecting our modified equations, we can make an educated guess for a solution given by 
\begin{equation}
  \mathcal{G}^{rr} =  \hspace{.09cm}\hspace{.09cm}\mathcal{G}^{\theta \theta } = 0
\end{equation}
 It follows from the symmetry of the metric that 
\begin{equation}
  \mathcal{G}^{\phi\phi} =  r^2  \hspace{.09cm} sin^2\theta  \hspace{.09cm} \mathcal{G}^{\theta \theta } = 0
\end{equation}
Thus the modified equations reduce to a single equation given by

\begin{equation}
\mathcal{G}^{tt} \hspace{.09cm}  + {\dfrac{\sigma}{m^2}} \nabla_t \nabla_{t}(  g^{tt} \hspace{.09cm} \mathcal{G}^{tt} )-  {\dfrac{2 \sigma}{m^2}}\nabla_b \nabla_{a}(\dfrac 12 g^{ab} \hspace{.09cm} \mathcal{G}^{tt})
\end{equation}
\begin{equation}
- \nabla_t \nabla_{t}( \mathcal{G}^{tt} \hspace{.09cm} {\dfrac{4 \zeta}{m^2}}  \hspace{.09cm}(a^{t})^2) +\nabla_c \nabla_{t}( \mathcal{G}^{tt} \hspace{.09cm} {\dfrac{4 \zeta}{m^2}}  \hspace{.09cm}a^{c} a^{t} ) \nonumber
\end{equation}

\begin{equation}
 {+ \nabla_t \nabla_{d}( \mathcal{G}^{tt} \hspace{.09cm} {\dfrac{4 \zeta}{m^2}}  \hspace{.09cm}a^{t} a^{d}) -\nabla_c \nabla_{d}( \mathcal{G}^{tt} \hspace{.09cm} {\dfrac{4 \zeta}{m^2}}  \hspace{.09cm}a^{c} a^{d})  = 0 }
\end{equation}

Thus, it is clear that the solution to the modified Einstein's equation depends upon the polarization of the photon.In fact, this turns out to be the most important feature of the new gravity theory. To obtain a nice analytic solution and for the sake of simplicity, we look at a case where $ a^{r} =(i {m^2})/{4 \zeta}  $ and  $ a^{\theta } = a^{\phi } = 0 $. Thus, the equation becomes

\begin{eqnarray}
  \mathcal{G}^{tt} +  {\dfrac{\sigma}{m^2}} \nabla_t \nabla_{t}(  g^{tt} \hspace{.09cm} \mathcal{G}^{tt} )-  {\dfrac{2 \sigma}{m^2}}\nabla_b \nabla_{a}(\dfrac 12 g^{ab} \hspace{.09cm} \mathcal{G}^{tt}) \nonumber \\
+ \nabla_r \nabla_{r}( \mathcal{G}^{tt} \hspace{.09cm}  {\dfrac{m^2}{4 \zeta}}  )= 0
\end{eqnarray}

Calculating the covariant derivative w.r.t $ g_{ab}$, and neglecting product of infinitesimal terms, we obtain
\begin{equation}
  \mathcal{G}^{tt} +\bigl{[}\dfrac{m^2}{4 \zeta}- {\dfrac{\sigma}{m^2}}\bigl{]}\hspace{.09cm}\dfrac{d^2 \mathcal{G}^{tt}}{d r^2}  -  {\dfrac{2 \sigma}{m^2}}\nabla_\theta \nabla_{\theta}( \dfrac{ \mathcal{G}^{tt}}{r^2} ) = 0
\end{equation}
This can be further simplified to get :
\begin{equation}
 \mathcal{G}^{tt} +\bigl{[}\dfrac{m^2}{4 \zeta}- {\dfrac{\sigma}{m^2}}\bigl{]}\hspace{.09cm} \dfrac{d^2 \mathcal{G}^{tt}}{d r^2}  -  {\dfrac{2 \sigma}{m^2}}\bigl{[} \dfrac{1}{r} \dfrac{d\mathcal{G}^{tt}}{d r} - \dfrac{2}{r^3} \dfrac{d \mathcal{G}^{tt}}{d r} +\dfrac{6 {G}^{tt}}{ r^4}  \bigl{]}= 0 \label{eq:ten}
\end{equation}

However, these equations are too complicated to have a well-behaved analytic solution. But, the equation can be reduced to a simpler form if we make some sensible approximations. Plugging in the values of the $\zeta , \sigma \hspace{.09cm}\text{and} \hspace{.09cm} m$, we find that  
\begin{equation}
  \dfrac{2\sigma}{m^2} \sim  10^{-32}  \bigl{[}\dfrac{m^2}{4 \zeta}- {\dfrac{\sigma}{m^2}}\bigl{]} \hspace{.09cm} \sim 10^{-16} 
\end{equation}

Thus, for very large $r$, the contribution of the third term in square bracket of Eq.~(\ref{eq:ten}) is negligible and we get the following equation :
\begin{equation}
 \mathcal{G}^{tt} +\bigl{[}\dfrac{m^2}{4 \zeta}- {\dfrac{\sigma}{m^2}}\bigl{]}\hspace{.09cm} \dfrac{d^2 \mathcal{G}^{tt}}{d r^2} = 0 \label{soln}
\end{equation}

This is an ordinary differential equation in one variable which can be solved easily. A physical solution is 
\begin{equation}
\mathcal{G}^{tt} =  e^{- k r} ,  \hspace{.09cm} \text{where} \hspace{.09cm}   k  = \frac{1}{\sqrt{\bigl{[}\dfrac{-m^2}{4 \zeta}+ {\dfrac{\sigma}{m^2}}\bigl{]}} }
\end{equation}
 Here  $ k$ is a constant and as $\zeta$ is negative, $k$ is real and positive. It is imperative that we check the consistency of the solution obtained before proceeding any further. We can easily see that only at very large $r$,  $\mathcal{G}^{tt}$ turns out to be infinitesimal in magnitude and this is necessary for our effective metric $G_{ab}$ to  satisfy the  Eq.~(\ref{metric})

This condition renders the contribution of the third term of Eq.~(\ref{eq:ten}) negligible and we end up getting  Eq.~(\ref{soln})

 Starting from the effective metric  $ G_{ab}$, we can calculate the components of the Einstein tensors given by  
\begin{equation}
  \mathcal{G}^{tt} = \dfrac{-w'}{r} +\dfrac{h''}{r^2} -\dfrac{h'}{2r^3} -\dfrac{3 h }{2 r^4}- \dfrac{w}{r^2} 
\end{equation}
\begin{equation}
  \mathcal{G}^{rr} = \dfrac{-v'}{r} -\dfrac{h'}{2r^3} -\dfrac{3 h }{2 r^4}+ \dfrac{w}{r^2} 
\end{equation}

\begin{equation}
 \mathcal{G}^{ \theta \theta} =-\dfrac{v''}{2 r^2} -\dfrac{(v'-w')}{2r^3} +\dfrac{3 h }{2 r^6}+\dfrac{h'}{2r^5} 
\end{equation}
\begin{equation}
  \mathcal{G}^{\phi\phi} =  r^2  \hspace{.09cm} sin^2\theta  \hspace{.09cm} \mathcal{G}^{\theta \theta } 
\end{equation}

Since the last three equations vanish, we can freely set $h(r)=0$ as $w(r) = r v'(r)$ solves both the second and third equation. Thus, first equation implies
\begin{equation}
  \dfrac{-w'}{r} - \dfrac{w}{r^2} =  e^{- k r}
\end{equation}
 This can be easily solved to get the solution

\begin{equation}
  w(r) = \dfrac{f}{r} + \dfrac{e^{-k r} \hspace{.09cm}\left(k^2 \hspace{.09cm}r^2+2\hspace{.09cm} k \hspace{.09cm}r+2\right)}{k^3\hspace{.09cm} r}
\end{equation}
 The constant of integration can be fixed by taking the classical limit of $r \to  \infty$ and identifying the leading order term of the metric component as that of the Schwarzschild metric for a very large $ r $ ,viz. $(1 + \dfrac{2  \hspace{.09cm} G \hspace{.09cm}M}{r})$. Thus, we get
\begin{equation}
  w(r) = \dfrac{2  \hspace{.09cm} G \hspace{.09cm}M}{r} + \dfrac{e^{-k r} \hspace{.09cm}\left(k^2 \hspace{.09cm}r^2+2\hspace{.09cm} k \hspace{.09cm}r+2\right)}{k^3\hspace{.09cm} r}
\end{equation}
 From $w(r) = r v'(r)$ , we get 
\begin{equation}
 v(r) = - \dfrac{2  \hspace{.09cm} G \hspace{.09cm}M}{r}  -\dfrac{e^{-k r} \hspace{.09cm} (k \hspace{.09cm} r+2)}{k^3 \hspace{.09cm} r}
\end{equation}
 Thus, we observe that the photon ``sees" a modified Schwarzschild metric, with a very small quantum correction. However, there is a peculiar property to this solution which is absent in the Schwarzschild solution of general relativity. Let us solve for the radius at which the radial component of velocity of the photon vanishes. This radial distance corresponds to the event horizon of the black hole. Since the photon travels along null geodesics in geometric optics approximation, we have

\begin{equation}
 G_{ab} \hspace{.09cm} u^{a}\hspace{.09cm} u^{b} = \mathcal{B}(r) \hspace{.09cm} \dot t^2+ \mathcal{A}(r) \hspace{.09cm} \dot r^2 =0
\end{equation}

 Here, $ u^{a}$ is the tangent vector along the trajectory of the photon, parametrized by some $\tau$ and the dot is taken w.r.t $\tau$ . The expression translates to
\begin{equation}
  \dfrac{dr}{dt} = \sqrt{- \mathcal{B}(r)/\mathcal{A}(r)}
\end{equation}
\begin{equation}
 ~~ ~~~~~~~~~~~~  = \sqrt{(1+v(r)) (1-w(r))} \label{hor}
\end{equation}
 The LHS is the radial component of the co-ordinate velocity of the photon. The dependence of the velocity with $r$ can analyzed using Eq.~(\ref{hor}). Plugging in the values of the functions, we can easily see that the radial component vanishes not at the Schwarzschild radius, but at a slightly closer value of $r$. This implies that the event horizon depends upon the direction of polarization of the light.

Since the metric $G_{ab}$ is lorentzian, the predictability of the matter equations is guaranteed and the principle polynomial becomes bi-hyperbolic. Thus, the solution doesn't violate the initial assumptions of the theory.

Now let us consider a photon which is not polarized in any particular direction. Thus, the components $a^{k}$ will be varying in both negate and positive directions. As the frequency of this variation is greater than the reaction rate of the metric to the fluctuations of the electromagnetic field, we take a statistical average and find that $\langle  a^{r} \rangle  \hspace{.09cm}=  \hspace{.09cm}\langle a^{t} \rangle \hspace{.09cm} =  \hspace{.09cm}\langle  a^{\theta} \rangle \hspace{.09cm} = \langle  a^{\phi} \rangle = 0$. Thus, if we set  $ \mathcal{G}^{rr} =  \hspace{.09cm}\hspace{.09cm}\mathcal{G}^{\theta \theta } = 0$, the new equations reduce to the Einstein's equations. Hence, we get back the Schwarzschild solution.

  This is an interesting phenomenon as some photons ``see" the event horizon at a particular distance while other photons with different polarization see the horizon at another radial co-ordinate.

\section{Discussion }
 We have studied how the interaction of photons and the metric modifies the Maxwell's equation in QED and how this leads to new gravitational dynamics. The alteration of local photon propagation has lead to the modification of the Maxwell's equation. For these new matter equations to have a well-posed Cauchy problem, it is necessary for the underlying gravitational field to have a modified dynamics. These constraints lead to a new modified theory of gravity.

 The most surprising feature of the theory is that the modified Einstein's equation is dependent on the polarization of the photon. This is analogous to the dependence of wavelength of light in the rainbow gravity theory[4]. However, it should be noted that when we were formulating the theory, we started off with just electromagnetic field. We did not include other forms of matter to determine the Lagrangian for gravity. Light cones of the photons will cease to dictate the trajectories of massive fields upon the inclusion of other such matter fields. This is because massive fields couple to the metric through their fields or the components of the fields in some representation space of their symmetry group and may behave non-trivially.

It should also be noted that only kinematics of photons of very long wavelengths are described by the theory as the modified equations are derived using geometric optics ansatz. Thus, it is necessary to account for higher energy  photons in developing a better approximation to this modified theory .

 Another consequence of the modifications is that for a homogeneous and isotropic universe, the dependence of polarization of the metric translates to the fact that if we trace out the past of each photon, we end up getting different singularities in spacetime. This has a direct implication that a particle sees the universe older or younger than another particle with a different polarization. Thinking along the lines of Ref.\cite{smolin04}, we find that this could solve the Horizon problem or even takes us further away from a solution. 

  The event horizon of a black hole becomes dependent on the polarization of the photon. The exact location of horizon of the black hole is different for different polarization of light. Thus, we end up getting an effective event horizon rather than an absolute event horizon. This dependence of polarization might give us some clues in solving the information loss paradox \cite{PhysRevD.14.2460} and improve our understanding of black holes.

\section{Acknowledgments }
I am very grateful to Aninda Sinha and Chethan Krishnan of Center for High Energy Physics, Indian Institute of Science for their useful and insightful comments on the paper.

\nocite{*}

\bibliography{apssamp}

\end{document}